\documentclass[runningheads]{llncs}
\usepackage{graphicx}
\usepackage{hyperref}
\usepackage{xcolor}
\usepackage{amsmath}  
\usepackage{wrapfig,lipsum,booktabs}
\usepackage{multirow} 
\usepackage{wrapfig}
\usepackage[ruled,norelsize]{algorithm}
\usepackage{algpseudocode,algorithmicx}
\usepackage{textcomp}
\usepackage{framed}
\usepackage[T1]{fontenc}
\usepackage{mathptmx}

\definecolor{darkblue}{rgb}{0,0,0.6}
\newcommand{\blt}[1]{\textcolor{black}{#1}}
\newcommand{\bluetexttt}[1]{\textcolor{black}{\texttt{#1}}}
\newcommand{\relup}[1]{\textcolor{blue}{+#1\%}}
\newcommand{\reldown}[1]{\textcolor{red}{-#1\%}}
\newcommand{\NameQR}{GenQREnsemble}
\newcommand{\NamePRF}{GenQREnsembleRF}
\newcommand{\bluet}[1]{\textcolor{darkblue}{#1}}

\begin{document}%

\title{\NameQR{}: Zero-Shot LLM Ensemble Prompting for Generative Query Reformulation}
\titlerunning{\NameQR{} and \NamePRF{}}

\author{Kaustubh D. Dhole,
Eugene Agichtein} 
\institute{Department of Computer Science\\
Emory University \\
Atlanta, USA \\
\email{\{kaustubh.dhole, eugene.agichtein\}@emory.edu}}

%
\maketitle              
%
\begin{abstract}
Query Reformulation(QR) is a set of techniques used to transform a user’s original search query to a text that better aligns with the user’s intent and improves their search experience. Recently, zero-shot QR has been shown to be a promising approach due to its ability to exploit knowledge inherent in large language models. By taking inspiration from the success of ensemble prompting strategies which have benefited many tasks, we investigate if they can help improve query reformulation. In this context, we propose an ensemble based prompting technique, \NameQR{} which leverages paraphrases of a zero-shot instruction to generate multiple sets of keywords ultimately improving retrieval performance. We further introduce its post-retrieval variant, \NamePRF{} to incorporate pseudo relevant feedback. On evaluations over four IR benchmarks, we find that \NameQR{} generates better reformulations with relative nDCG@10 improvements up to 18\% and MAP improvements upto 24\% over the previous zero-shot state-of-art. On the MSMarco Passage Ranking task, \NamePRF{} shows relative gains of 5\% MRR using pseudo-relevance feedback, and 9\% nDCG@10 using relevant feedback documents.
\end{abstract}
\section{Introduction}
Users searching for relevant documents might not always be able to accurately express their information needs in their initial queries. This could result in queries being vague or ambiguous or lacking the necessary domain vocabulary. Query Reformulation (QR) is a set of techniques used to transform a user’s original search query to a text that better aligns with the user’s intent and improves their search experience. Such reformulation alleviates the vocabulary mismatch problem by expanding the query with related terms or paraphrasing it into a suitable form by incorporating additional context.

Recently, with the success of large language models (LLMs)~\cite{gpt3,i2}, a plethora of QR approaches have been developed. The generative capabilities of LLMs have been exploited to produce novel queries~\cite{doc2query}, as well as useful keywords to be appended to the users' original queries~\cite{zeroshot}. By gaining exposure to enormous amounts of text during pre-training, prompting has become a promising avenue for utilizing knowledge inherent in an LLM for the benefit of subsequent downstream tasks~\cite{bigbench} especially QR~\cite{promptingqueryexpa,whenfail}. 

Unlike training or few-shot learning, zero-shot prompting does not rely on any labeled examples. The advantage of a zero-shot approach is the ease with which a standalone generative model can be used to reformulate queries by prompting a templated piece of instruction along with the original query. Particularly, zero-shot QR can be used to generate keywords by prompting the user's original query along with an instruction that defines the task of query reformulation in natural language like~\bluetexttt{Generate useful search terms for the given query:`List all the breweries in Austin'}. 

\begin{wrapfigure}{r}{0.5\textwidth}
\centering
        \resizebox{0.5\textwidth}{!}{%
    \begin{tabular}{ll}
         \textbf{Instruction} & \textbf{Expansions Generated}\\ \hline
         Increase the search efficacy by offering & age  goldfish  grow  outsmart  outlive \\
         beneficial expansion keywords for the query & ageing  species...\\ \hline
         Enhance search outcomes by recommending beneficial & Goldfish breed sizes What kind of\\ 
         expansion terms to supplement the query & goldfish grows the fastest...\\ \hline \\
    \end{tabular}}
    \caption{\blt{Keywords generated for the query (``do goldfish grow'') differ drastically when generated from two paraphrastic instructions prompted to~\bluetexttt{flan-t5-xxl}~\cite{flant5}.}}
    \label{tab:example}
\end{wrapfigure}

However, such a zero-shot prompting approach is still contingent on the exact instruction appearing in the prompt providing plenty of avenues of improvement. While LLMs have been known to vary significantly in performance across different prompts~\cite{robust1,robust2} and generation settings~\cite{decoding}, many natural language tasks have benefited by exploiting such variation via ensembling multiple prompts or generating diverse reasoning paths~\cite{diverse,ama,selfcons}. Whether such improvements also transfer to tasks like QR is yet to be determined. In Figure~\ref{tab:example}, a vast difference is noticed in the keywords generated when the input instruction is altered to a semantically similar variant. We hypothesize that QR might naturally benefit from such variation -- An ensemble of zero-shot reformulators with paraphrastic instructions can be tasked to look at the input query in diverse ways so as to elicit different expansions. This work proposes the following contributions:
\begin{itemize}
    \item We propose a novel method,\textbf{\NameQR{}} -- a zero-shot \textbf{Ensemble} based \textbf{Gen}erative  \textbf{Q}uery \textbf{R}eformulator which exploits multiple zero-shot instructions for QR to generate a more effective query reformulation than possible with an individual instruction. (Section 3)
    \item We further introduce an extension \textbf{\NamePRF{}} to incorporate \textbf{R}elevance \textbf{F}eedback into the process. (Section 3)
    \item We evaluate the proposed methods over four standard IR benchmarks, demonstrating significant relative improvements vs. recent state of the art, of up to 18\% on nDCG@10 in pre-retrieval settings, and of up to 9\% nDCG@10 on post-retrieval (feedback) settings, demonstrating increased generalizability of our approach.
\end{itemize}
Next, we summarize the prior work to place our contributions in context.

\section{Related Work}
Query reformulation has been shown to be effective in many settings~\cite{qrsurvey}. It can be done pre-retrieval, or post-retrieval, via incorporating evidence from feedback, obtained either from a user or from top-ranked results in the sparse retrieval setting~\cite{prfsurvey}, and in the dense retrieval setting~\cite{colbertprf,anceprf}. 

Recently, zero-shot approaches to query reformulation have received considerable attention. Wang et al.~\cite{zeroshot} design a query reformulator by fine-tuning a sequence-to-sequence transformer, T5~\cite{t5} on pairs of raw and transformed queries. Their zero-shot prompting approach uses an instruction-tuned model, FlanT5~\cite{flant5} to generate keywords for query expansion and incorporating PRF. Jagerman et al.~\cite{promptingqueryexpa} demonstrate LLMs can be more powerful than traditional methods for query expansion. Mo et al.~\cite{convsearch} propose a framework to reformulate conversational search queries using LLMs. Gao et al.~\cite{hyde}'s framework performs retrieval through fake documents generated by prompting LLMs with user queries. Alaofi et al.~\cite{queryvariants} prompt LLMs with information descriptions to generate query variants.

However, using a single query reformulation can sometimes degrade performance compared to the original query. To address this drawback, prior efforts have incorporated ensemble strategies via keywords from numerous sources or fusing documents from different queries. Gao et al.~\cite{qu2}, combine features derived from various translation models to generate better query rewrites. Si et al.~\cite{qu6} perform QR by utilizing multiple external biomedical resources. Hsu and Taksa~\cite{qu7} present a data fusion framework suggesting that diverse query formulations represent distinct evidence sources for inferring document relevance. Later, Mohankumar et al.~\cite{qu1} generated diverse queries by introducing a diversity-driven RL algorithm. For other tasks, recent works demonstrated the benefits of ensemble strategies for prompting LLMs, including self-consistency~\cite{selfcons} for arithmetic and common sense tasks, Chain of Verification~\cite{cove} for improving factuality, and Diverse~\cite{diverse} for question answering. However, zero-shot based ensemble methods for LLM have not been explored for the Query Reformulation task, as we propose in this paper.

\section{Proposed Approach: \NameQR{}}

In this section, we describe two variations of our proposed approach, for the pre- and post-retrieval settings. In the pre-retrieval setting, a Query Reformulation $R$ transforms a user's expressed query $q_0$ into a novel reformulated version $q_r$ to improve retrieval effectiveness for a given search task (e.g., passage or document retrieval). 
We also consider the post-retrieval setting, wherein the reformulator can incorporate additional contextual information like document or passage-level feedback. 
\label{sub2}
\begin{figure*}
  \includegraphics[width=\textwidth]{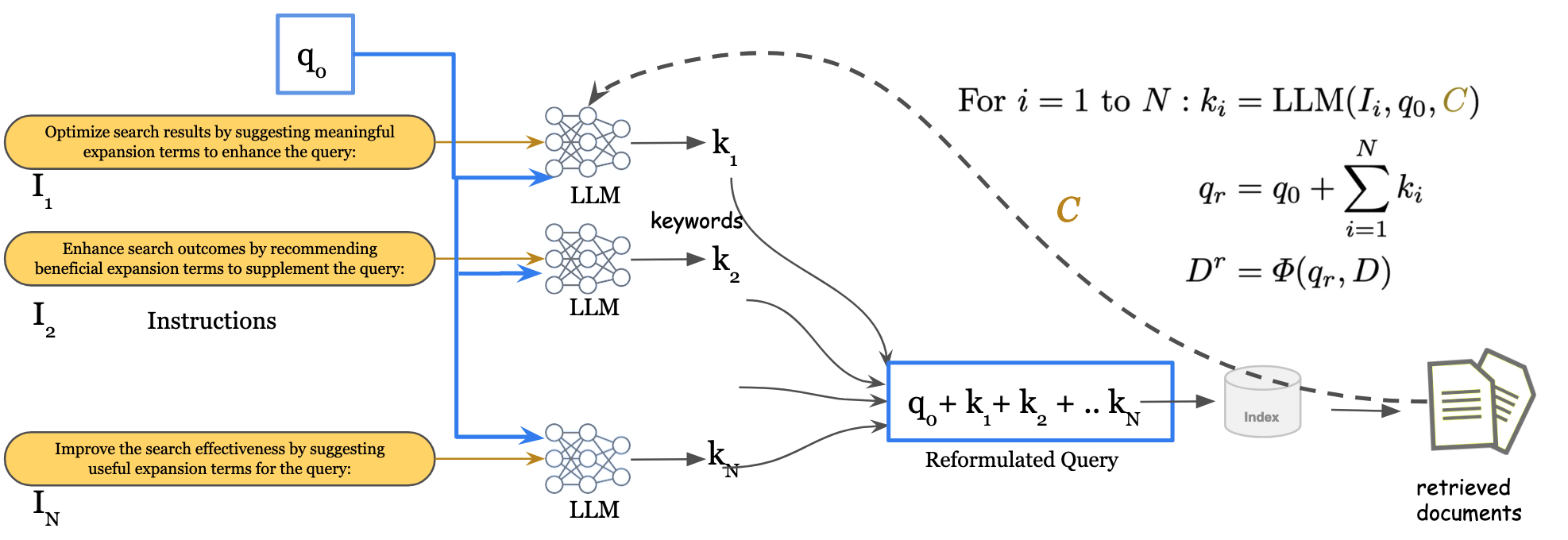}
  \caption{\blt{The complete flow and algorithm shown on the top right.}}
  \label{fig:unifiedquery}
\end{figure*}

\textbf{Pre-retrieval}: We propose \textbf{\NameQR{}} -- an ensemble prompting based query reformulator which uses $N$ diverse paraphrases of a QR instruction to enhance retrieval. Specifically, we first use an LLM to paraphrase the instruction $I_1$ to create $N$ instructions with different surface forms viz. $I_1$ to $I_{N}$. This is required to be done once. Each instruction is then prompted along with the user's query $q_0$ to generate instruction-specific keywords. All the keywords are then appended to the original query, resulting in a reformulated query, which is then executed against a document index $D$ to retrieve relevant documents $D^{r}$. The complete process and algorithm are shown in Figure~\ref{fig:unifiedquery}.

\textbf{Post-retrieval}: To assess how well our method can incorporate additional context like document feedback, we introduce \textbf{\NamePRF{}}. Here, we prepend the $N$ instructions described earlier with a fixed context capturing string ``\bluetexttt{Based on the given context information \{C\},}'' used\footnote{We found prepending the string in the prompt performs better than appending it at the end} in ~\cite{zeroshot} to create their PRF counterparts -- where \bluetexttt{C} is a space (` ') delimited concatenation of feedback documents \bluetexttt{C} $= d_1 + \ldots + d_m$, obtained either as pseudo-relevance feedback from initial retrieval or manually chosen by the user.
\section{Experiments}\label{experiments}

\begin{wrapfigure}{r}{0.5\textwidth}
\centering
\label{tab1}
\resizebox{0.5\textwidth}{!}{%
\begin{tabular}{ll}
\# &  Instruction\\
\hline
1 &  \bluetexttt{Improve the search effectiveness by suggesting expansion terms for the query}\\
2 &  \bluetexttt{Recommend expansion terms for the query to improve search results}\\
3 &  \bluetexttt{Improve the search effectiveness by suggesting useful expansion terms for the query}\\
4 &  \bluetexttt{Maximize search utility by suggesting relevant expansion phrases for the query}\\
5 &  \bluetexttt{Enhance search efficiency by proposing valuable terms to expand the query}\\
6 &  \bluetexttt{Elevate search performance by recommending relevant expansion phrases for the query}\\
7 &  \bluetexttt{Boost the search accuracy by providing helpful expansion terms to enrich the query}\\
8 &  \bluetexttt{Increase the search efficacy by offering beneficial expansion keywords for the query}\\
9 &  \bluetexttt{Optimize search results by suggesting meaningful expansion terms to enhance the query}\\
10 &  \bluetexttt{Enhance search outcomes by recommending beneficial expansion terms to supplement the query}\\ \\
\end{tabular}
}
\caption{\blt{Reformulation instructions generated ($N$=10).}}
\end{wrapfigure}

We now describe the experiments and analysis performed for different retrieval settings. 

To instruct the LLM to generate query reformulations, we start with the instruction empirically chosen by Wang et al.~\cite{zeroshot} -- as our base QR instruction $I_1$. We use this instruction to generate $N$ paraphrases of the instruction ($N=10$). To this aim, we invoke \href{https://chat.openai.com/}{GPT-3.5} API with the paraphrase generating prompt, namely, $I_p$=\bluetexttt{Generate 10 paraphrases for the following instruction:}-- and the base QR instruction $I_1$ to obtain $I_2$ to $I_{10}$. These paraphrases serve as our instruction set for subsequent experiments.

For generating the actual query reformulations, we employ~\bluetexttt{flan-t5-xxl}~\cite{flant5}, an instruction-tuned model. The FlanT5 set of models is created by fine-tuning the text-to-text transformer model,~\bluetexttt{T5}~\cite{t5} on instruction data of a variety of NL tasks. We use the checkpoint\footnote{\url{https://huggingface.co/google/flan-t5-xxl}} provided through HuggingFace’s Transformers library~\cite{huggingface}. Nucleus sampling is performed with a cutoff probability of 0.92 keeping the top 200 tokens (top\_k) and a repetition penalty of 1.2.

For evaluation, we use four popular benchmarks through IRDataset~\cite{irds}'s interface: 1)\textbf{TP19}: TREC 19 Passage Ranking which uses the MSMarco dataset~\cite{msmarco,promptingqueryexpa} consisting of search engine queries. 2)\textbf{TR04}: TREC Robust 2004 Track, a task intended for testing poorly performing topics. In our experiments, we use the Title as our choice of query. And two tasks from the BEIR~\cite{beir} benchmark 3)\textbf{WT}: Webis Touche~\cite{touche2020} for argument retrieval 4)\textbf{DE}: DBPedia Entity Retrieval~\cite{dbpedia}.
\subsection{Baselines:} 
We compare our work against the following using the Pyterrier~\cite{pyterrier} framework. For all the post-retrieval experiments, we use 5 documents as feedback.\\
\textbf{With BM25 Retriever}:
\begin{itemize}
    \item BM25: Here, we retrieve using raw queries without any reformulation
    \item FlanQR~\cite{zeroshot}: We implement Wang et al's single-instruction zero-shot QR~\cite{zeroshot} which is also a specific case of \NameQR{}  when N=1
    \item BM25+RM3~\cite{rm3}: BM25 retrieval with RM3 expanded queries (\#feedback terms=10)
    \item BM25+FlanPRF~\cite{zeroshot}: BM25 retrieval with FlanPRF expanded queries
\end{itemize}
\textbf{With Neural Reranking}: Here, we re-evaluate the above settings in conjunction with a MonoT5 neural reranker~\cite{monoT5} with all other parameters constant.
\begin{itemize}
    \item BM25+MonoT5: BM25 retrieval using raw queries, re-ranked with MonoT5 model~\cite{monoT5}  
    \item FlanQR+MonoT5: BM25 retrieval with FlanQR reformulations, re-ranked with MonoT5 model
    \item BM25+RM3+MonoT5: BM25 retrieval with RM3 expanded queries, re-ranked with MonoT5 model
    \item BM25+FlanPRF+MonoT5: BM25 retrieval with FlanPRF expanded queries, re-ranked with MonoT5 model
\end{itemize}

\section{Results and Analysis}
We now report the results of query reformulation for pre- and post-retrieval settings.
\subsection{Pre-Retrieval Performance}
\begin{wrapfigure}{r}{0.5\textwidth}
\vspace{-40pt}
\resizebox{0.5\textwidth}{!}{%
\centering
\includegraphics[width=0.45\textwidth]{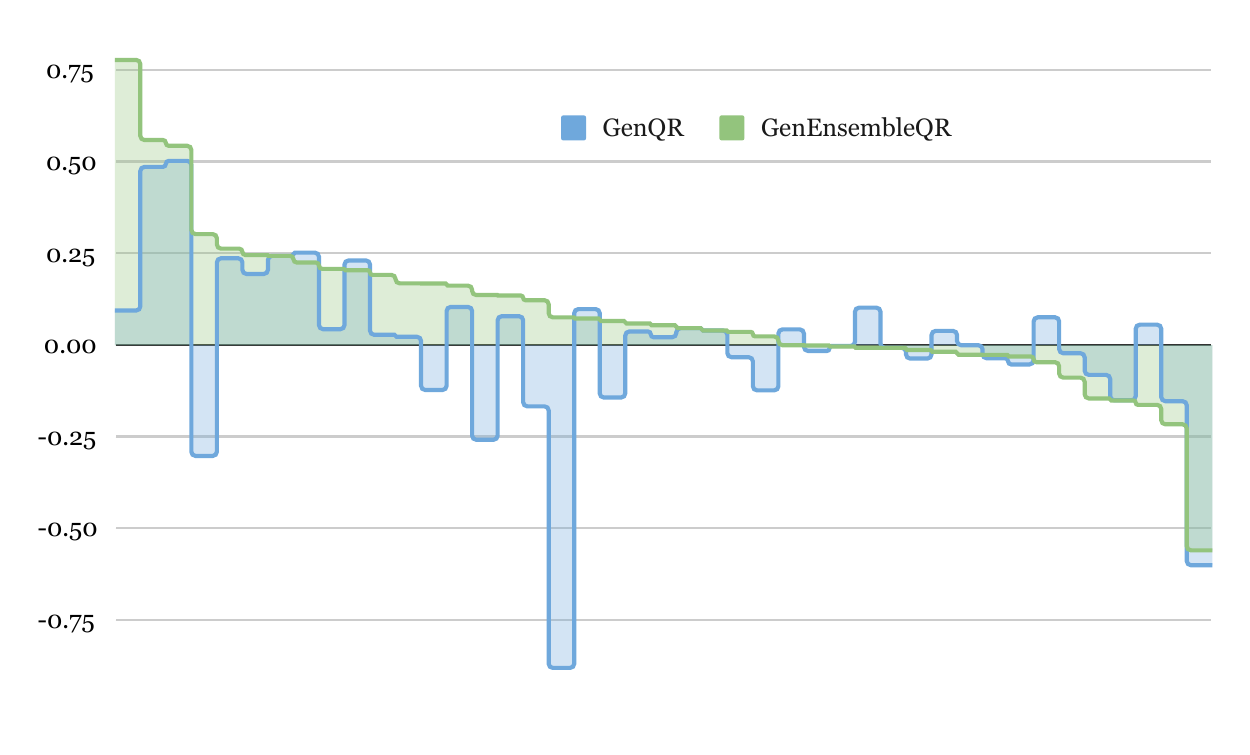}}
\vspace{-30pt}
\caption{\blt{nDCG@10 Scores of \NameQR{} and FlanQR  relative to BM25}}
\vspace{-15pt}
\label{querywise}
\end{wrapfigure}
We first compare the retrieval performances of raw queries and reformulations from FlanQR, and \NameQR{} in Table~\ref{ensembleperformance}.\NameQR{}  outperforms the raw queries as well as generates better reformulations than FlanQR's reformulated queries across all the four benchmarks over a BM25 retriever, indicating the usefulness of paraphrasing initial instructions. On TP19, nDCG@10 and MAP improve significantly with relative improvements of 18\% and 24\% respectively. This is further validated through the querywise analysis shown in Figure~\ref{querywise} -- Relative to BM25, nDCG@10 scores of \NameQR{} (shown in green) are overall better than FlanQR (shown in blue).\NameQR{} seems more robust too as it avoids drastic degradation in at least 6 queries on which FlanQR fails.

\begin{table}[!ht]
\centering
\caption{\blt{Performance of \NameQR{} on the four benchmarks. $\alpha$ denotes significant improvements (paired t-test with Holm-Bonferroni correction, $p<0.05$) over FlanQR. \relup{} indicates \% improvements relative to FlanQR (as whole numbers).}}
    \resizebox{\textwidth}{!}{ %
    \begin{tabular}{llll|lll|lll|lll}
     Evaluation Set &
      \multicolumn{3}{c}{ \textbf{TREC Passage 19} } &
      \multicolumn{3}{c}{ \textbf{TREC Robust 04} } &
      \multicolumn{3}{c}{ \textbf{Webis Touche} } &
      \multicolumn{3}{c}{ \textbf{DBpedia Entity} } \\ \hline
        \textbf{Setting} & nDCG@10 & MAP & MRR & P@10 & nDCG@10 & MRR & nDCG@10 & MAP & MRR & nDCG@10 & MAP & MRR \\ 
        BM25 & .480 & .286 & .642 & .426 & .434 & .154 & .260 & .206 & .454 & .321 & .168 & .297 \\ 
        FlanQR & .477 & .302 & .593 & .473 & .483 & .151 & .315 & .241 & .511 & .342 & .196 & .345 \\ 
        FlanQR$_{\beta=.05}$ & .511 & .323 & .621 & .469 & .477 & .150 & .276 & .221 & .476 & .353 & .188 & .339 \\
        \bluet{\NameQR{}} & \textbf{.564}$^\alpha$\relup{18} & \textbf{.375}$^{\alpha}$\relup{24} & \textbf{.706}\relup{19} & \textbf{.500}$^\alpha$\relup{6} & \textbf{.513}$^\alpha$\relup{6} & \textbf{.159}\relup{6} & \textbf{.317}\relup{1} & \textbf{.257}\relup{6} & \textbf{.555}\relup{9} & \textbf{.374}$^\alpha$\relup{9} & \textbf{.212}$^\alpha$\relup{8} & \textbf{.376}$^\alpha$\relup{9} \\ 
        \bluet{\NameQR{}$_{\beta=.05}$} & \textbf{.575}$^\alpha$ & \textbf{.377}$^\alpha$ & \textbf{.714} & \textbf{.502}$^\alpha$ & \textbf{.512}$^\alpha$ & \textbf{.159} & .292 & \textbf{.242} & .489 & \textbf{.377}$^\alpha$ & \textbf{.212}$^\alpha$ & \textbf{.380}$^\alpha$ \\ \hline
        BM25+MonoT5 & .718 & .477 & \textbf{.881} & \textbf{.492} & \textbf{.513} & \textbf{.173} & \textbf{.299} & .216 & .525 & .414 & .249 & .444 \\ 
        FlanQR+MonoT5 & .707 & .486 & .847 & .490 & .510 & .170 & .292 & .215 & .530 & .415 & .255 & .446 \\ 
        \bluet{\NameQR{}+MonoT5} & \textbf{.722}\relup{2} & \textbf{.503}\relup{3} & .862\relup{2} & .484\reldown{1} & .506\reldown{1} & .170 & .298\relup{3} & \textbf{.219}\relup{2} & \textbf{.548}\relup{3} & \textbf{.420}\relup{1} & \textbf{.258}\relup{1} & \textbf{.450}\relup{1} \\  \\
  \end{tabular}}
\label{ensembleperformance}
\end{table}

We further look at \NameQR{}  under the neural reranker setting shown at the bottom half of Table~\ref{ensembleperformance}. In three of the four settings, viz., TP19, WT, and DE, \NameQR{} is preferable to its zero-shot variant, FlanQR. Evidently, the gains of both the zero-shot approaches in the traditional setting are stronger vis-à-vis the neural setting. We hypothesize this could be due to \NameQR{} and FlanQR both expanding the query via incorporating semantically similar but lexically different keywords. Comparatively, neural models are adept at capturing notions of semantic similarity and might benefit less with query expansion. This also is in line with Weller et al.'s~\cite{whenfail} recent analysis on the non-ensemble variant.
\vspace{-7pt}
\subsection{Post-Retrieval Performance}
\begin{table}[!ht]
\caption{\blt{Comparison of PRF performance on the TREC 19 Passage Ranking Task}}
    \centering
    \resizebox{\textwidth}{!}{%
    \begin{tabular}{lllll|llll}
      &
      \multicolumn{4}{c}{ \textbf{With BM25 Retriever} } &
      \multicolumn{4}{c}{ \textbf{With Neural Reranking} } \\ \hline
        Setting & nDCG@10 & nDCG@20 & MAP & MRR & nDCG@10 & nDCG@20 & MAP & MRR \\
        BM25 & .480 & .473 & .286 & .642 & .718 & .696 & .477 & .881 \\ 
        RM3 & .504 & .496 & .311 & .595 & .716 & .699 & .480 & .858 \\ 
        FlanPRF & .576 & .553 & .363 & .715 & .722 & .703 & .486 & .874 \\ 
        \bluet{\NamePRF{}} & \textbf{.585}\relup{2} & \textbf{.560}\relup{1} & \textbf{.373}\relup{3} & \textbf{.753}\relup{5} & \textbf{.729}\relup{1} & \textbf{.706}\relup{1} & \textbf{.501}\relup{3} & \textbf{.894}\relup{2} \\ \hline 
        FlanPRF (Oracle) & .753 & .728 & .501 & .936 & .742 & .734 & .545 & .881 \\ 
        \bluet{\NamePRF{}} (Oracle) & \textbf{.820}$^{\alpha}$\relup{9} & \textbf{.773}\relup{6} & \textbf{.545}\relup{9} & \textbf{.977}\relup{4} & \textbf{.756}\relup{2} & \textbf{.751}\relup{2} & \textbf{.545} & \textbf{.897}\relup{2} \\
    \end{tabular}}
    \label{prf_results}
    \vspace{-20pt}
\end{table}
We now investigate if \NamePRF{} can effectively incorporate PRF in Table~\ref{prf_results}. We find that \NamePRF{} improves retrieval performance as compared to other PRF approaches and is able to incorporate feedback from a BM25 retriever better than RM3 as well as its zero-shot counterpart. To assess if \NamePRF{} and FlanPRF can at all benefit from incorporating relevant documents, we perform oracle testing by providing the highest relevant gold documents as context. We find that \NamePRF{} is able to improve over \NameQR{} (without feedback) showing that it is able to capture context effectively as well as benefit from it. Further, it can incorporate relevant feedback better than its single-instruction counterpart FlanPRF. We notice improvements even under the neural reranker setting as \NamePRF{} outperforms RM3 and FlanPRF. Besides, the oracle improvements are higher with only a BM25 retriever as compared to when a neural reranker is introduced. \\
\begin{figure}
\centering
\begin{minipage}{0.5\textwidth}
  \centering
  \includegraphics[width=\linewidth]{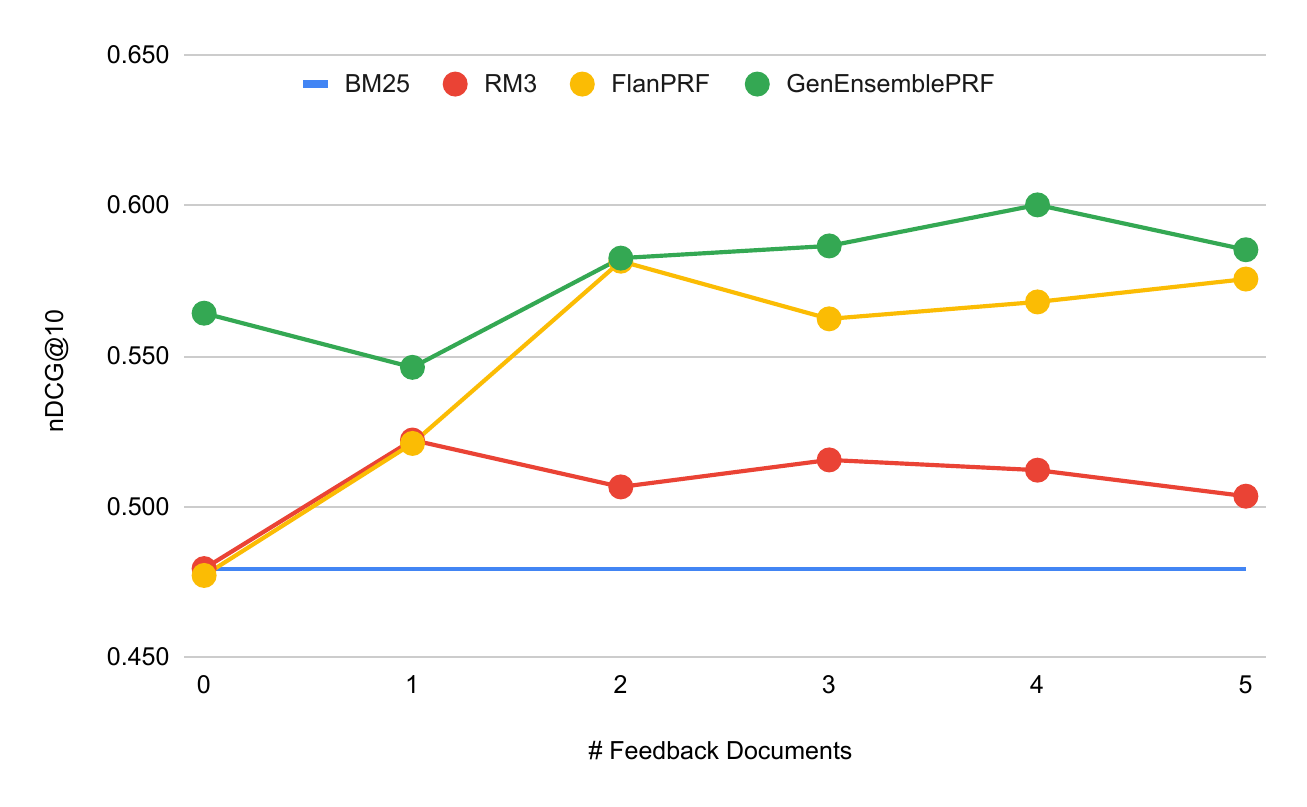}
  \label{fig:test1}
\end{minipage}%
\begin{minipage}{0.5\textwidth}
  \centering
  \includegraphics[width=\linewidth]{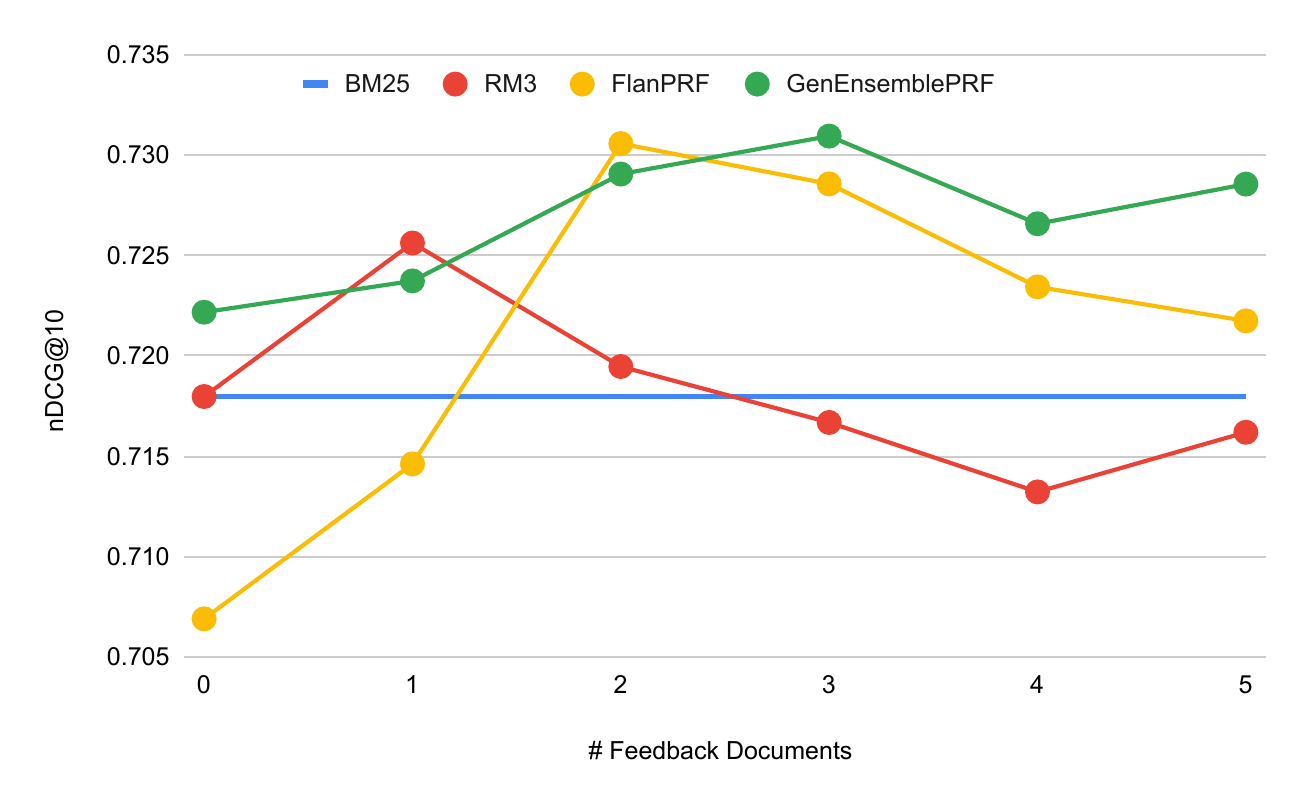}
  \label{fig:test2}
\end{minipage}
\caption{\blt{Effect of feedback documents under sparse (BM25) and neural (MonoT5) rankers}}
\end{figure}

We further evaluate the effect of varying the number of feedback documents from 0 to 5. We notice that resorting to an ensemble approach is highly beneficial. In the BM25 setting, the ensemble approach seems always preferable. Under the neural reranker setting too,\NamePRF{} almost always outperforms FlanPRF.

\section{Conclusions}
Zero-shot QR is advantageous since it does not rely on any labeled relevance judgements and allows eliciting pre-trained knowledge in the form of keywords by prompting the model with the original query and appropriate instruction. By introducing \NameQR{}, we show that zero-shot performance can be further enhanced by using multiple views of the initial instruction. We also show that the extension \NamePRF{} is able to effectively incorporate relevance feedback, either automated or from users. While generative QR greatly benefits from our ensemble approach, the proposed methods come at a cost of potentially increased latency, but this is becoming less problematic with the increased availability of batch inference for LLMs. The proposed ensemble approach could also be applied to other settings, for example, to address different aspects of queries or metrics to optimize, or to better control the generated reformulations. 

\bibliographystyle{splncs04}
\bibliography{bibtext}

\end{document}